\documentclass[final,3p,times,twocolumn]{elsarticle}  

\usepackage{amssymb}
\usepackage{amsmath}
\usepackage{amsthm}
\usepackage[english]{babel}
\usepackage{xspace}
\usepackage{cases}
\usepackage{color}
\usepackage{graphicx}
\usepackage[font=footnotesize]{caption}
\usepackage[font=footnotesize]{subcaption}
\usepackage{epstopdf}
\usepackage{upgreek}
\usepackage{booktabs}
\usepackage{array}
\usepackage{blindtext}
\usepackage{multirow}
\usepackage{bbold}
\usepackage[hidelinks]{hyperref}
\usepackage{soul}
\usepackage{enumerate}
\usepackage{blindtext}
\usepackage{graphicx}
\usepackage{xcolor}
\usepackage{siunitx}
\usepackage{pgf}
\usepackage{lipsum}
\usepackage[normalem]{ulem}
\usepackage{glossaries}

\newcommand{\figwidth}{\linewidth}

\newcommand{\nbus}{N}

\newcommand{\nline}{N_l}

\newcommand{\horizon}{T}
\newcommand{\pnet}{p}
\newcommand{\gen}{u}
\newcommand{\load}{d}

\newcommand{\storage}{s}
\newcommand{\energy}{e}
\newcommand{\lineflow}{c}

\newcommand{\caseNoStorage}{\textsc{S}1\xspace}
\newcommand{\caseStorage}{\textsc{S}2\xspace}
\newcommand{\caseStorageWithVariance}{\textsc{S}3\xspace}

\newcommand{\Load}{\MakeUppercase{\load}}
\newcommand{\Gen}{\MakeUppercase{\gen}}
\newcommand{\Storage}{\MakeUppercase{\storage}}

\newcommand{\ptdfmat}{\Phi}

\newcommand{\adjustlength}{-4.5mm}

\newcommand{\opf}{\textsc{opf}\xspace}

\newcommand{\dc}{\textsc{dc}\xspace}

\newcommand{\socp}{\textsc{socp}\xspace}

\newcommand{\rv}[1]{\mathsf{#1}}

\newcommand{\ev}[1]{\mathbb{E}(#1)}

\newcommand{\var}[1]{\mathbb{V}(#1)}

\newcommand{\prob}[1]{\mathbb{P}( #1 )}

\newcommand{\pvar}{u}
\newcommand{\pvarup}{\expandafter\uppercase\expandafter{\pvar}}

\newcommand{\pfix}{d}
\newcommand{\pfixup}{\expandafter\uppercase\expandafter{\pfix}}

\definecolor{till}{rgb}{.1,.4,.9}
\definecolor{timm}{rgb}{1.0,.0,1.0} 
\definecolor{veit}{rgb}{0.,.8,.4}
\definecolor{revised}{rgb}{.2,.8,.1}

\newacronym{ess}{ESS}{energy storage system}
\newacronym{cc}{CC}{chance-constrained}
\newacronym{opf}{OPF}{optimal power flow}
\newacronym{socp}{SOCP}{second-order cone problem}
\newacronym{res}{RES}{renewable energy source}
\newacronym[
  longplural={Gaussian processes}
]{gp}{GP}{Gaussian process}
\newacronym{gpr}{GPR}{Gaussian process regression}

\usepackage{hyperref}

\usepackage{cleveref}

\newcommand{\PaperTitle}{Analytical Uncertainty Propagation for Multi-Period Stochastic Optimal Power Flow}
\newcommand\scalemath[2]{\scalebox{#1}{\mbox{\ensuremath{\displaystyle #2}}}}

\journal{Journal of \LaTeX\ Templates}

\begin{document}

\begin{frontmatter}

\title{\PaperTitle}

\author[add1]{Rebecca Bauer}
\ead{rebecca.bauer@kit.edu}
\author[add1]{Tillmann~M\"uhlpfordt}
\ead{t.muehlpfordt@mailbox.org}
\author[add2]{Nicole Ludwig}
\ead{nicole.ludwig@uni-tuebingen.de}
\author[add1]{Veit Hagenmeyer}
\ead{veit.hagenmeyer@kit.edu}
\address[add1]{Institute for Automation und applied Informatics, Karlsruhe Institute of Technology}
\address[add2]{Cluster of Excellence “Machine Learning: New Perspectives for Science”, University of T\"ubingen, Germany}

\begin{abstract}

The increase in \glspl{res}, like wind or solar power, results in growing uncertainty also in transmission grids. 
This affects grid stability through fluctuating energy supply and an increased probability of overloaded lines. 
One key strategy to cope with this uncertainty is the use of distributed \glspl{ess}. 
In order to securely operate power systems containing renewables and use storage, optimization models are needed that both handle uncertainty and apply \glspl{ess}.
This paper introduces a compact dynamic stochastic chance-constrained DC optimal power flow~(CC-OPF) model, that minimizes generation costs and includes distributed \glspl{ess}. Assuming Gaussian uncertainty, we use affine policies to obtain a tractable, analytically exact reformulation as a \gls{socp}.
We test the new model on five different IEEE networks with varying sizes of 5, 39, 57, 118 and 300 nodes and include complexity analysis. 
The results show that the model is computationally efficient and robust with respect to constraint violation risk. The distributed energy storage system leads to more stable operation with flattened generation profiles. Storage absorbed RES uncertainty, and reduced generation cost.

\end{abstract}


\end{frontmatter}



\section{Introduction}

\glsresetall

Volatile renewable energy resources, such as wind, are increasingly included in power systems. Besides many benefits, these \glspl{res} bring more uncertainty into the system as they depend on fluctuating weather dynamics. This challenges the grid's reliability and leads to frequency fluctuations or RES curtailment. To cope with these new challenges, more and more research focuses on the operation of power systems under uncertainty \cite{Capitanescu12,Vrakopoulou17,Roald18,Guo18,Li17,Warrington13}. 
A central strategy to securely operate power systems under especially short-term (daily) uncertainty is the inclusion of distributed \glspl{ess} to mitigate the uncertainty. We focus on battery storage, which is good at short-term storage, but not long-term. 
E.g., currently, many grid boosters (large battery storage units) are installed in transmission grids \cite{molina_distributed_2012}; in Germany \cite{figgener_development_2021}, Europe \cite{zsiboracs_intermittent_2019}, and the world \cite{gulagi_role_2018, keck_impact_2019}. 
In contrast to conventional power plants (e.g., thermal, gas), \glspl{ess} have the advantage of costing less, and they can store and discharge power of renewables. During the thermal power plants' ongoing lifetime, they can also react much quicker to fluctuations. 


In order to tackle the short-term uncertainty in power systems together with storage, we can use DC \gls{opf}. \Gls{opf} is used to minimize costs in a power system while respecting both the physical limits of the network such as line flow limits, and the power flow equations. It is a standard tool in network operation (e.g., optimal generation scheduling) and the DC linearization of AC power flow is a standard approximation method \cite{wood_power_2014}. 
Under stochastic uncertainty the DC OPF can be formulated as a chance-constrained OPF (CC-OPF), which is exact when assuming Gaussian uncertainty \cite{Muehlpfordt18c}. 

According to \cite{Warrington13}, any method including uncertainty should encompass three important aspects:
\begin{enumerate}
\itemsep0em 
	\item \label{item:forecasts} forecasts of uncertain disturbances such as feed-in from solar and wind sources, or demand;
	\item \label{item:policies} control policies for reacting to errors in forecasts;
	\item \label{item:propagation} propagation of uncertainties over time and space.
\end{enumerate}
Literature yet combines only aspects \ref{item:policies} and \ref{item:propagation}. Before we come to that, we list some applications for each aspect individually:
For the first aspect of forecasts of uncertain disturbances current literature proposes various methods that predict entire distributions or quantiles, see e.g. \cite{Hong16} for an overview, for different renewable energy sources \cite{quatile_zhang_2020, quantile_solar_2017}. 
For the second aspect of the control policies, affine control policies are often applied to problems related to the operation of power systems under uncertainty. These applications range from reserve planning \cite{Warrington13,Ding16,Bucher17} and dispatching in active distribution networks \cite{Fabietti17,Fabietti18}, to optimal power flow \cite{Vrakopoulou13b,Vrakopoulou17,Muehlpfordt18c,Louca16,MunozAlvarez14}, or building control \cite{Oldewurtel12}. 
Affine policies are already in use in power systems, and convince with their simple structure and their ability to split the power balance equation nicely such that it is fulfilled despite uncertainties~\cite{Muehlpfordt18c}.
For the third aspect of the propagation of uncertainty, efficient methods have been proposed. For example, scenario-based approaches \cite{Capitanescu12,Fabietti17,Vrakopoulou17}, and approaches employing polynomial chaos \cite{Muehlpfordt17a,Muehlpfordt16b,Muehlpfordt18c}. Other works study multi-period (propagation over time) \gls{opf} under uncertainty, but employ scenario-based or data-driven approaches~\cite{Vrakopoulou13,Vrakopoulou17b,Capitanescu12,Guo18}.

Approaches combining both affine policies and propagation over time and space are to be found in both robust and stochastic optimization. 
In general, robust optimization does not assume an underlying uncertainty distribution, but an uncertainty set of parameters. Hence, since there is no parametrization of the input distribution, there is no exact reformulation of the equations. An exact reformulation of a stochastic optimization problem is possible, since the input assumes one exact uncertainty distribution. 
A concept in between robust optimization and stochastic optimization is distributionally robust optimization (DRO). DRO assumes an ambiguity set containing distributions, hence, an exact reformulation is also not possible.
In stochastic optimization, on the other hand, there are several approaches for uncertainty propagation.
For example, there are several multi-period \glspl{opf} with storage that assume Gaussian uncertainty, however, they often do not include CCs or affine policies. They use scenario-trees \cite{hemmati_stochastic_2016}, others look at AC power flow in a distribution network \cite{ayyagari_chance_2017}, or approximate the \glspl{cc} \cite{summers_stochastic_2015, li_analytical_2015}. While some works do offer an exact reformulation of CCs, they are either static \cite{bienstock_chance_2012}, lack storages \cite{bienstock_chance_2012, roald_analytical_2013}, or do not include affine policies \cite{sjodin_risk-mitigated_2012}.
Few approaches offer models including \glspl{cc} and a formulation into a \gls{socp}, but lack affine policies and time \cite{zhang_distributionally_2017}, look at the risk of cost functions without storage \cite{xie_distributionally_2018}, or apply different chance constraints \cite{Warrington13}. 
The latter approaches differ to the methodology introduced in the present paper, and often also in their problem objective. Also, many of them focus on detailed modelling of specific parts, while we hold our formulation general.

The contribution of this paper, contrary to the above, combines it all: Firstly, we include all three aspects; forecasts, control policies and uncertainty propagation. Secondly, we include chance constraints into a multi-period second-order cone problem formulation, adding storage. And lastly, our stochastic optimal power flow model is computationally efficient for mid-size networks and an analytically exact equivalent of the original formulation.

Specifically, we optimize generation and storage schedules using a stochastic chance-constrained multi-period \gls{opf} that is subject to time-varying uncertainties and contains a large energy storage system. 
We choose to use \glspl{gp} to describe the uncertain demand and generation, as they are well-suited to model power time series \cite{mitrentsis_probabilistic_2021}. \Glspl{gp} are very flexible~\cite{Roberts12} and allow a closed-form expressions of random variables, since they consist of Gaussian distributions that stay Gaussian when passed through some linear operator (such as the linear DC OPF). This assumption allows an exact reformulation of the problem.
The idea of an "analytical reformulation" has been used in \cite{Li17}, however, they focus on joint chance constraints. Several works have applied \glspl{gp} to wind power forecasting~\cite{Kou13,Chen14}, solar power forecasting~\cite{Sheng18}, and electric load forecasting~\cite{Mori08,Leith04,Lloyd14,Rogers11,Blum13,McLoughlin13}. Given our modelling choice of \glspl{gp}, the natural way to forecast uncertain disturbances for different time horizons, e.g. a day, is through \gls{gpr} \cite{Rasmussen06} as it yields the desired mean and covariance for \glspl{gp}. 
Additionally, we use different risk levels for the chance constraints -- not to be confused with the risk of the cost function \cite{hemmati_stochastic_2016}.

To the best of our knowledge there are no works that model a DC multi-period CC-OPF, with affine policies and Gaussian uncertainty, in a transmission network, that is reformulated into a tractable, analytically exact equivalent, convex \gls{socp} \emph{and} including forecast of uncertainties via \gls{gpr}. In contrast to most literature we extensively test our model on various network sizes from 5 to 300 nodes.

The remainder of the paper is structured as follows. \Cref{sec:modelling_all} states the grid, models uncertainties as  \glspl{gp}, and introduces affine policies. \Cref{sec:OPF} states the yet intractable optimization problem under uncertainty. Then, \Cref{sec:SolutionMethodology} reformulates the \opf problem as a tractable convex optimization problem, and comments on its computational characteristics. The case studies in \Cref{sec:CaseStudy} apply the proposed optimization approach to the \textsc{ieee} 5-bus, 39-bus, 57-bus, 118-bus, and 300-bus test cases and a complexity analysis is provided. Lastly, the results are discussed in \Cref{sec:discussion}.

\section{Modelling assumptions}
\label{sec:modelling_all}

The model of the electrical grid is at the core of the optimization. 
Let us consider a connected graph with $\nbus$ buses and $\nline$ lines 
under \dc power flow conditions 
for time instants $\mathcal{\horizon} = \{ 1, \dots, \horizon \}$.
Every bus $i \in \mathcal{\nbus} =  \{1, \dots, \nbus \}$ can contain a disturbance $\load_i(t), i\in\mathcal{\Load}\subseteq\mathcal{\nbus}$, (i.e., load or renewables), a thermal generation unit $\gen_i(t)$, $i\in\mathcal{\Gen}\subseteq\mathcal{\nbus}$, and a storage injection unit~$\storage_i(t)$, $i\in\mathcal{\Storage}\subseteq\mathcal{\nbus}$.

We denote the power excess/deficit at node $i$ and time $t$ as
\begin{equation}
    \pnet_i(t) = \load_i(t) + \gen_i(t) + \storage_i(t),
\end{equation}
which is also the total power export/influx into/from node $i$ needed to guard the nodal power balance \cite{horsch_linear_2017}. 

In the following we will model the uncertain disturbances, as well as generation and storage that react to the disturbance and are modelled accordingly.

\subsection{Uncertain Disturbances as Gaussian Processes}
\label{sec:GP}

Uncertain disturbances are loads and volatile feed-ins from renewable energies. We denote them by~$\load_i(t)$ at bus~$i \in \mathcal{\nbus}$ and time~$t \in \mathcal{\horizon}$.
Specifically, we assume in this paper that the uncertainty is Gaussian and that the disturbances have no spatial correlation, i.e. the random variables are independent. For wind in particular Gaussianity of the prediction error is reasonable through the central limit theorem, since a large set of agglomerated wind farms has a normally distributed power output \cite{hemmati_stochastic_2016}.  This uncertain disturbance
is the realization of a discrete-time stochastic process $\{ \rv{\load}_i(t) \: \forall t \in \mathcal{\horizon} \}$ given by\,\footnote{More precisely, $\{ \rv{\load}_i(t) \: \forall t \in \mathcal{\horizon} \}$ is only a snapshot of the overarching stochastic process $\{ \rv{\load}_i(t) \: \forall t \in \tilde{\mathcal{\horizon}} \}$, where $\tilde{\mathcal{\horizon}}$ is an infinite set, and $\mathcal{\horizon} \subset \tilde{\mathcal{\horizon}}$ a finite subset thereof.
We however neglect this subtlety for the sake of simplicity in the present paper.}
\begin{subequations}
	\label{eq:ConditionOnStochasticProcess}
	\begin{align}
	\label{eq:ConditionOnStochasticProcess_Mapping}
	\scalemath{0.92}{%
		\begin{bmatrix}
		\rv\load_i(1) \\
		\rv\load_i(2) \\
		\vdots \\
		\rv\load_i(\horizon) \\
		\end{bmatrix}
		= \underbrace{\begin{bmatrix}
			[\hat\load_i]_1\\
			[\hat\load_i]_2 \\
			\vdots \\
			[\hat\load_i]_{\horizon}
			\end{bmatrix}}_{=: \hat\load_i} +
		\underbrace{\begin{bmatrix}
			[\Load_i]_{11} & 0 & \hdots & 0 \\
			[\Load_i]_{21} & [\Load_i]_{22} & & 0 \\
			\vdots & & \ddots\\
			[\Load_i]_{\horizon 1} & & & [\Load_i]_{\horizon \horizon}]
			\end{bmatrix}}_{=: \Load_i} 
		\underbrace{\begin{bmatrix}
			[\rv\Xi_i]_1 \\
			[\rv\Xi_i]_2 \\
			\vdots \\
			[\rv\Xi_i]_\horizon		
			\end{bmatrix}}_{
			=: \rv\Xi_i}
	} 
	\end{align}
for all buses $i \in \mathcal{\nbus}$,
where $\hat\load_i \in \mathbb{R}^{\horizon}$ is the mean vector and $\Load_i \in \mathbb{R}^{\horizon \times \horizon}$ the lower-triangular, non-singular covariance matrix.
The stochastic germ~$\rv\Xi_i$ is a $\horizon$-variate Gaussian random vector whose elements are independent Gaussian random variables $[\rv\Xi_i]_j\sim\mathcal{N}(0,1)$\,\footnote{Notice that non-singularity of $\Load_i$ means that~\eqref{eq:ConditionOnStochasticProcess_Mapping} is a one-to-one mapping between $[\rv\load_i(1), \dots,	\rv\load_i(\horizon)]^\top$ and the stochastic germ~$\rv\Xi_i$. The lower-triangularity of $\Load_i$ allows to create this mapping first for time instant $t=1$, then $t=2$, etc.\label{foot:OnetoOneMapping}}.
Hence, the forecast error is Gaussian.
The lower-triangularity of $\Load_i$ means that the uncertain disturbance~$\rv\load_i(t)$ is causal, i.e. 
\begin{align}
\label{eq:UncertaintyModel}
\rv\load_i(t) = [ \hat\load_i ]_t + \sum_{k=1}^{t} [\Load_i]_{tk} [\rv\Xi_i]_k,
\end{align}
where $\rv\load_i(t)$ depends only on past and present time instants $k = 1, \dots, t$, but not on future ones.
\end{subequations}
Every uncertain disturbance is then fully described by its mean~$\ev{\rv\load_i(t)}$ and variance~$\var{\rv\load_i(t)}$, which we need to provide for the given time horizon
\begin{equation}
\label{eq:MeanVariance_Load}
\ev{\rv\load_i(t)} = [ \hat\load_i ]_t, \quad
\var{\rv\load_i(t)} = \sum_{k=1}^{t} [\Load_i]_{tk}^2.
\end{equation}

\subsection{Affine Policies}

Having parametrized the uncertain disturbances in an affine fashion, the reaction of generation and storage is modelled accordingly. In particular, the latter have to to assume uncertainty themselves as uncertainty means that they can react to errors in forecasts. Otherwise, the power balance equation could not be fulfilled. Therefore, we model generation and storage analogously to the uncertainty: as realizations of (affine) random processes $\{ \rv\gen_i(t) \, \forall t \in \mathcal{\horizon} \}$ and $\{ \rv\storage_i(t) \, \forall t \in \mathcal{\horizon} \}$, respectively.

We do that by introducing \textit{affine policies} that determine how generation and storage react to the uncertain disturbances. For generation we introduce feedback of the form
\begin{subequations}
\label{eq:GenerationPolicy}
\begin{align}
\label{eq:AffineFeedback_Gen}
\rv\gen_i
= \hat{\gen}_i + \sum_{j \in \mathcal{\nbus}} \Gen_{i,j} \rv\Xi_j, \quad \forall i \in \mathcal{\nbus},
\end{align}
\end{subequations}
for all time instants $t \in \mathcal{\horizon}$\,\footnote{Notice that the control policy~\eqref{eq:AffineFeedback_Gen} is written in terms of the stochastic germs~$\rv\Xi_j$ for $j \in \mathcal{\nbus}$;
but in practice it is the realization of the uncertain disturbances $\rv\load_i(t)$ that can be measured.
It is always possible to get the realization of the stochastic germ from the realization of the uncertain disturbance, and vice versa, see Footnote~\ref{foot:OnetoOneMapping}.\label{foot:Rewriting}}.
For this, $\hat{\gen}_i \in \mathbb{R}^{\horizon}$, and every $\Gen_{i,j} \in \mathbb{R}^{\horizon \times \horizon}$ with $j \in \mathcal{\nbus}$ is lower-triangular. 
The latter enforces the feedback to be causal, as they cannot depend on future uncertainties.
Note that the notation is structurally equivalent to~\eqref{eq:ConditionOnStochasticProcess} with the same stochastic germ.

We introduce the same kind of feedback policy~\eqref{eq:AffineFeedback_Gen} for the storage injections (from storage to grid)
\begin{align}
\label{eq:StoragePolicyPerBusPerTime}
\rv\storage_i
= \hat{\storage}_i + \sum_{j \in \mathcal{\nbus}} \Storage_{i,j} \rv\Xi_j,
\end{align}
where $\hat{\storage}_i\in\mathcal{R}^{\horizon}$ and every $\Storage_{i,j} \in \mathbb{R}^{\horizon \times \horizon}$ with $j \in \mathcal{\nbus}$ is lower-triangular.

Having established $\load_i(t)$, $\gen_i(t)$ and $\storage_i(t)$ we can further derive closed-form expressions of the other random variables.
From storage injections $\rv\storage_i(t)$ we can directly model the actual storage states $\rv\energy_i(t)$ as discrete-time integrators
\begin{align}
\label{eq:StorageDynamics_RV}
\rv\energy_i(t+1) = \rv\energy_i(t) - h \, \rv\storage_i(t), \quad \rv\energy_i(1) = \rv\energy_i^{\textsc{ic}} \quad \forall i \in \mathcal{\nbus}.
\end{align}
Reformulating the equation towards $\rv\storage_i(t)$, $\, \rv\storage_i(t) = \frac{\rv\energy_i(t)-\rv\energy_i(t+1)}{h}$, makes clear that $\rv\storage_i(t)$ ($-\rv\storage_i(t)$) is the discharge (charge) of storage from time $t$ to $t+1$, or the injection into the network.
In general, uncertainty also affects the initial condition~$\rv\energy_i^{\textsc{ic}}$ of storage~$i$.
For simplicity, firstly, the value of $h > 0$ subsumes the discretization time. Theoretically, it could also be used as a potential loss factor. A second simplification is that we do not use different variables and efficiencies for charging and discharging as in \cite{powermodels_julia}.

Moreover, the change of generation inputs can be derived as $\Delta\rv\gen_i(\tau) = \rv\gen_i(\tau) {-} \rv\gen_i(\tau {-} 1)$ and the net power becomes $\rv\pnet_i(t) = \rv\load_i(t) + \rv\gen_i(t) + \rv\storage_i(t)$ for bus $i$. 
Lastly, using the power transfer distribution matrix $\ptdfmat$ mapping net power to line flows, the line flow can be expressed as $\rv\lineflow_j(t) = \ptdfmat_{j} [\rv\pnet_1(t), \dots, \rv\pnet_{\nbus}(t)]^\top$. 
The voltage angles are implicitly contained in the definition of the net power $\rv\pnet_i(t)$\,\footnote{The Kirchhoff Current Law (KCL) is given by the net power formula $\pnet_i = \sum_lK_{il}f_i$  $\forall i=1,\dots,N$, where the line flows are $f_l = \frac{1}{x_l}\sum_iK_{il}\phi_i$ $\forall l=1,\dots,N_l$ with voltage angles $\phi_i$ and incidence matrix $K$ \cite{horsch_linear_2017}}.
Note that all those random variables are  \glspl{gp} by linearity. Hence, as such they are fully described by their mean and variance, as listed in Table~\ref{tab:Moments}.

\begin{table*}
	\centering
	\caption{Closed-form expressions for state of storage~$\rv\energy_i(t+1)$, change of inputs~$\Delta\rv\gen_i(\tau)$, and line flows~$\rv\lineflow_j(t)$.\label{tab:FunctionalDepenendencies}}
	\small
	\begin{tabular}{lll}
		\toprule
		$ \rv\lineflow_l(t) $ & $=$ & $  \displaystyle\sum_{i \in \mathcal{\nbus}} [\ptdfmat]_{li} ( [\hat\load_i]_t + [ \hat\gen_i ]_t +[ \hat\storage_i ]_t ) + \displaystyle\sum_{i \in \mathcal{\nbus}} \sum_{k=1}^{t} \Big( [\ptdfmat]_{li} [\Load_i]_{tk} + \displaystyle \sum_{j \in \mathcal{\nbus}} [\ptdfmat]_{lj} ( [\Gen_{j,i}]_{tk} + [\Storage_{j,i}]_{tk} ) \Big) [\rv\Xi_i]_k $ \\
		$\Delta\rv\gen_i(\tau)$ & $=$ & $ [\hat\gen_i]_{\tau} - [\hat\gen_i]_{\tau {-} 1} + \displaystyle \sum_{j \in \mathcal{\nbus}} \Big( [\Gen_{i,j}]_{\tau \tau} [\rv\Xi_j]_{\tau} + \displaystyle \sum_{k=1}^{\tau {-} 1} \Big( [\Gen_{i,j}]_{\tau k} - [\Gen_{i,j}]_{(\tau {-} 1) k} \Big) [\rv\Xi_j]_k \Big)$ \\
		$ \rv\energy_i(t+1)$ & $=$ & $\rv\energy_i^{\textsc{ic}} - h \displaystyle \sum_{k=1}^t [\hat\storage_i]_k - h \displaystyle \sum_{j \in \mathcal{\nbus}} \sum_{k=1}^t \Big( \displaystyle \sum_{l=k}^t [\Storage_{i,j}]_{lk} \Big) [\rv\Xi_j]_k $ \\
		\bottomrule
	\end{tabular}
\end{table*}
\begin{table*}
	\centering
	\caption{Expected value and variance of random variables from Problem~\eqref{eq:CCOPF_original} under affine policies \eqref{eq:GenerationPolicy} and \eqref{eq:StoragePolicyPerBusPerTime}.\label{tab:Moments}}
	\small
	\begin{tabular}{lll|lll}
		\toprule
		$\rv x$ & $\ev{\rv x}$ & $\var{\rv x} = \sigma^2$ & $\rv x$ & $\ev{\rv x}$ & $\var{\rv x} = \sigma^2$\\
		\midrule
		$\rv\load_i(t)$ & $[\hat\load_i]_t$ & $\displaystyle\sum_{k=1}^{t} [\Load_i]_{tk}^2$ & $\rv\lineflow_l(t)$ & $\displaystyle\sum_{i \in \mathcal{\nbus}} [\ptdfmat]_{li} ( [\hat\load_i]_t + [ \hat\gen_i ]_t +[ \hat\storage_i ]_t )$ & $ \displaystyle\sum_{i \in \mathcal{\nbus}} \displaystyle \sum_{k=1}^{t} \Big(  [\ptdfmat]_{li} [\Load_i]_{tk} + \displaystyle \sum_{j \in \mathcal{\nbus}} [\ptdfmat]_{l,j} ([\Gen_{j,i}]_{tk} + ([\Storage_{j,i}]_{tk}) \Big)^2$\\
		$\rv\gen_i(t)$ & $[ \hat\gen_i ]_t$ & $\displaystyle\sum_{j \in \mathcal{\nbus}} \displaystyle\sum_{k=1}^{t} [ \Gen_{i,j} ]_{tk}^2$ & $\Delta\rv\gen_i(\tau)$ & $[ \hat\gen_i ]_{\tau} - [ \hat\gen_i ]_{\tau{-}1}$ &  $\displaystyle\sum_{i \in \mathcal{\nbus}} \Big( [\Gen_{i,j}]_{\tau \tau}^2 {+} \sum_{k=1}^{\tau{-}1} ( [\Gen_{i,j}]_{\tau k} - [\Gen_{i,j}]_{(\tau-1) k})^2 \Big) $\\
		$\rv\storage_i(t)$ & $[ \hat\storage_i ]_t$ & $\displaystyle\sum_{j \in \mathcal{\nbus}} \displaystyle\sum_{k=1}^{t} [ \Storage_{i,j} ]_{tk}^2$ & $\rv\energy_i(t+1)$ & $\ev{\rv\energy_i^{\textsc{ic}}} - h \displaystyle \sum_{k=1}^{t} [ \hat\storage_i ]_k $	 & $\var{\rv\energy_i^{\textsc{ic}}} + h^2 \displaystyle \sum_{j \in \mathcal{\nbus}} \displaystyle \sum_{k=1}^{t} \Big(\displaystyle \sum_{l=k}^{t} [\Storage_{i,j}]_{lk} \Big)^2$ \\	
		\bottomrule
	\end{tabular}
\end{table*}

\subsection{Local and global balancing}
\label{sec:localglobal}

We have formulated the generators response to uncertainty through affine policies. Furthermore, we can specify how exactly generators react through the structure of the matrices ~$\Gen_{i,j}$, called \emph{local} and \emph{global balancing}.

In local balancing each generator $i$ reacts to every possible source of uncertainty $\rv{\Xi}_j$
Global balancing lets each generator react to the \emph{sum} of deviations and can be achieved by enforcing $\Gen_{i,1} = \hdots = \Gen_{i,\nbus}$ \cite{Muehlpfordt18c}.


%

\subsection{Predicting Uncertainties with Gaussian Process Regression}
\label{sec:GPR}
To predict the uncertain disturbances~$\rv\load_i$, we need the mean $\hat\load_i$ and covariance matrix $\Load_i$.
\gls{gpr} is a prediction method that yields precisely those.
GPR fits a family of functions $\mathcal{F}$ individually onto a data set $\mathcal{X}$. The posterior Gaussian process is then determined by the mean functions $\mu(t) = \mathbb{E}[\mathcal{F}(t)]$ ($t\in\mathbb{R}$) of $\mathcal{F}$ and a continuous covariance function $k(t,t')$\,\footnote{$k$ is also called a \textit{kernel} and should be a positive definite function.}
, $t,t'\in\mathbb{R}$, yielding $\Load_i$.
Thereby, $k$ reflects both the variance around $\mu(t)$ for some $t$, as well as the covariance of two values for $t,t'$. We write the Gaussian process as $\mathcal{N}(\mu,k)$.
Since both $\mu$ and $k$ are continuous ($t\in\mathbb{R}$), for the prediction we can simply extract the discrete vector $\mu(t)\stackrel{\wedge}{=}\hat\load_i(t)$ and matrix $\Load_i$ by inserting all future $t\in\horizon$ into $\mu(t)$ and $(t,t')\in\horizon\times\horizon$ into $k$.
Then the Gaussian process at node $i$ is written as
\begin{equation}
    d_i = \mathcal{N}(\hat\load_i,\Load_i^2) \quad \forall i \in \nbus.
\end{equation}
For the kernel function $k$ we use the sum of cosine and squared exponential (i.e. RBF) with an added constant function--yielding
\begin{equation}
\label{eq:kernel}
    k = k_{cosine} + k_{RBF} + k_{constant},
\end{equation}
with
\begin{align*}
    &k_{cosine}(x,x') & &= \sigma_{1}^2\cos\left(2\pi \sum_i \frac{(x-x')}{l_{1}}\right), \\
    &k_{RBF}(x,x')    & &= \sigma_{2}^2 \exp{\left(-\frac{(x-x')^2}{2l_{2}^2}\right)}, \\
    &k_{constant}(x,x') & &= \sigma_{3},
\end{align*}
where $\sigma_i$ is the variance and $l_i$ the lengthscale parameter. The variance determines the average distance of some $f\in\mathcal{F}$ to the mean function $\mu=\mathbb{E}[\mathcal{F}(x)]$; the lengthscale determines the length of the 'wiggles' in $f$ \cite{duvenaud_automatic_nodate}. This allows us to model periodicity as well as larger trends and smaller variations.

Having modelled all decision variables as random variables (and described how the uncertain disturbance are obtained), we can now put them all together into an optimization problem.
\vspace{1em}


\section{Optimization problem for power systems under uncertainty}
\label{sec:OPF}

Given a network and Gaussian decision variables, we can now introduce constraints and an objective in order to formulate the optimal power flow problem. Besides limits for line flows, storage injections, states and final states, generators and change of generation, a main constraint is the power balance equation
\begin{equation}
    \sum_{i \in \mathcal{\nbus}} \rv\pnet_i(t) = 0.
\end{equation}
Note that this is not the original nodal power balance equation, but the uninodal power balance, as $\pnet$ is the excess/deficit at node $i$. 
The leading objective can be formulated as: \textit{''How can we operate generators optimally in the presence of uncertainty?''} (given storage systems) and we thus formulate the chance-constrained \opf problem as
\newcommand{\adjustCCOPF}{-10mm}
\begin{subequations} 
	\label{eq:CCOPF_original}
	\begin{align}
	\label{eq:CCOPF_original_cost}
	\underset{\rv{\gen}_i(t), \rv{\storage}_i(t)}{\operatorname{min}}~  & \sum_{t \in \mathcal{\horizon}}\sum_{i \in \mathcal{\nbus}} \ev{ f_i(\rv\gen_i(t)} \quad \mathrm{s.t.}\\
	\label{eq:CCOPF_original_PowerBalance}
	& \hspace{\adjustCCOPF} \sum_{i \in \mathcal{\nbus}} \rv\load_i(t) + \rv\gen_i(t) + \rv\storage_i(t) = 0 \\
	\label{eq:CCOPF_original_StorageDynamics}
	& \hspace{\adjustCCOPF} 	\rv\energy_i(t+1) = \rv\energy_i(t) - h \, \rv\storage_i(t), ~ \rv\energy_i(1) = \rv\energy_i^{\textsc{ic}} \\
	\label{eq:CCOPF_CCs}
	& \hspace{\adjustCCOPF} \prob{ \rv x(t) \leq \overline{x} } \geq 1 - \varepsilon,
	~ \prob{ \rv x (t) \geq \underline{x} } \geq 1 - \varepsilon  \\
	\label{eq:CCOPF_original_StdConstraint}
	& \hspace{\adjustCCOPF} 0 \leq \sqrt{\var{\rv{x}}} \leq \sigma_{\overline{\rv{x}}} \\
	& \hspace{\adjustCCOPF} \forall \rv{x} {\in} \{ \rv\lineflow_j (t), \rv\gen_i(t), 
	\Delta\rv\gen_i(\tau),\! \rv\energy_i(t {+} 1), \rv\energy_i(T), \rv\storage_i(t) \} \\
	\nonumber
	& \hspace{\adjustCCOPF} \forall i \in \mathcal{\nbus}, \, t \in \mathcal{\horizon}, \, \tau \in \mathcal{\horizon} \setminus \{1\}, j \in \mathcal{L},
	\end{align}
\end{subequations}
where $\varepsilon \in [0,0.1]$ is the risk factor\,\footnote{It is straightforward to modify Problem~\eqref{eq:CCOPF_original} to consider time-varying and quantity-depending risk levels $\varepsilon$, e.g. use $\overline\varepsilon_{\lineflow_j}(t)$ to specify the risk level for satisfying the upper limit of line $j$ at time $t$.}.
Problem~\eqref{eq:CCOPF_original} minimizes the expected cost of generation over time~\eqref{eq:CCOPF_original_cost},
while satisfying the power balance~\eqref{eq:CCOPF_original_PowerBalance}
and the storage dynamics~\eqref{eq:CCOPF_original_StorageDynamics} in terms of random processes\,\footnote{For ease of presentation we assume the storage has already been installed and that their operation does not incur costs.}.

All engineering limits are formulated with chance constraints~\eqref{eq:CCOPF_CCs}: the probability that the line flow $\rv\lineflow_j(t)$, the generation $\rv\gen_i(t)$, the generation ramp $\Delta\rv\gen_i(\tau)$, the storage $\rv\storage_i(t)$, $\rv\energy_i(t)$ are below/above their upper/lower limits shall be greater than or equal to $1 - \varepsilon$.
We add chance constraints for the terminal state of the storage, $\rv\energy_i(T)$, to allow for the storage to be at a predefined level (with high probability) at the end of the horizon. The inequality constraint~\eqref{eq:CCOPF_original_StdConstraint} allows to restrict the standard deviation of all occurring random variables. The restriction enables to reduce the variation of certain generation units to be small. Note that this model can easily be adapted to power plants without ramp constraints by removing the respective equations.

\begin{figure}
    \centering
    \hspace{-1.5cm}
    \includegraphics[width=0.55\textwidth]{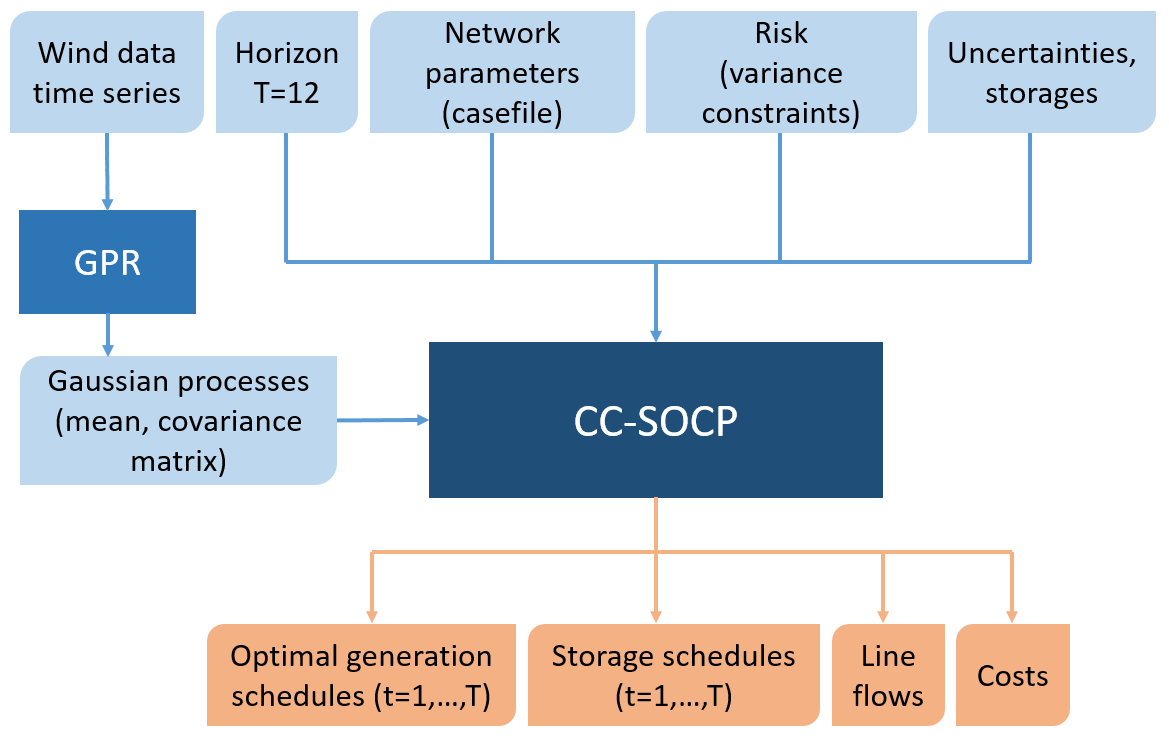}
    \caption{Inputs and outputs of the dynamic CC-SOCP.}
    \label{fig:model_input_output}
\end{figure}

Figure \ref{fig:model_input_output} visualizes this method, where the inputs are network parameters, uncertainties and storage, the time horizon, risk parameter, and predicted wind power as  \glspl{gp}. Then, the outputs are the optimal generation (decision variable) and its costs (objective), as well as storage schedules and line flows.


\section{Reformulation of Optimization Problem}
\label{sec:SolutionMethodology}

Problem~\eqref{eq:CCOPF_original} is intractable for several reasons: the decision variables are random processes, the equality constraints are infinite-dimensional, and the chance constraints and cost function require to evaluate integrals for the chance-constraints. In order to derive an exact yet finite-dimensional reformulation of the problem and cope with the intractability issues, we exploit the problems structure and the Gaussianity of all random variables. More specifically, we reformulate the infinite-dimensional power flow equation, compute the probabilities of the chance constraints, and rephrase the cost function.

\subsection{Power Balance}
\label{sec:PowerBalance}

To adapt the optimal power flow equations we start by taking the power balance~\eqref{eq:CCOPF_original_PowerBalance} and substituting both the uncertainty model~\eqref{eq:ConditionOnStochasticProcess} and the generation/storage control policies~\eqref{eq:GenerationPolicy}. Then, the power balance is satisfied for all realizations if \cite{Muehlpfordt18c}
\begin{subequations}
	\label{eq:PowerBalance_reformulated}
	\begin{align}
	\label{eq:PowerBalance_reformulated_mean}	
	\sum_{i \in \mathcal{\nbus}} \hat\load_i + \hat\gen_i + \hat\storage_i &= 0_{\horizon}, \\
	\label{eq:PowerBalance_reformulated_var}	
	\Load_j + \sum_{i \in \mathcal{\nbus}} \Gen_{i,j} + \Storage_{i,j} &= 0_{\horizon \times \horizon}, & \forall j \in \mathcal{\nbus}.
	\end{align}
\end{subequations}
Equation~\eqref{eq:PowerBalance_reformulated_mean} ensures power balance in the absence of uncertainties, or equivalently power balance in terms of the expected value; equation~\eqref{eq:PowerBalance_reformulated_var} balances every uncertainty~$\Load_j$ by the sum of the reactions from generation and storage.

\subsection{Chance Constraints}
\label{sec:ChanceConstraints}
As all random variables occurring in Problem~\eqref{eq:CCOPF_original} are Gaussian random variables, the chance constraints can be reformulated exactly using the first two moments:
Let $\rv x$ be a Gaussian random variable with mean $\mu$ and variance $\sigma^2$.
Then for $\varepsilon \in (0,0.1]$,
\begin{subequations}
	\label{eq:CC_Reformulation}
\begin{align}
&\prob{ \rv x \leq \overline{x}} \geq 1 - \varepsilon \quad \Longleftrightarrow & \hspace{-0.6cm} \phantom{\underline{x} \leq }\,\,\mu + \lambda(\varepsilon) \sqrt{\sigma^2} \leq \overline{x}, \\
&\prob{ \underline{x} \leq \rv x } \geq 1 - \varepsilon \quad  \Longleftrightarrow & \hspace{-0.6cm} \underline{x} \leq \mu - \lambda(\varepsilon) \sqrt{\sigma^2},
\end{align}
\end{subequations}
where $\lambda(\varepsilon) = \Psi^{-1}(1 {-} \varepsilon)$, and $\Psi$ is the cumulative distribution function of a standard Gaussian random variable \cite{Bienstock14}.
Hence, all chance constraints from Problem~\eqref{eq:CCOPF_original} can be reformulated by applying relation~\eqref{eq:CC_Reformulation} with the moments from Table~\ref{tab:Moments}.
Similarly, the constraint on the standard deviation~\eqref{eq:CCOPF_original_StdConstraint} is rewritten exactly using the expressions from Table~\ref{tab:Moments}.

\subsection{Cost Function}
\label{sec:CostFunction}
To rephrase the cost function, we consider quadratic generation costs
\begin{subequations}
\label{eq:Cost_Reformulation}
\begin{align}
f_i(\rv\gen_i(t)) = \gamma_{i,2} \rv\gen_i(t)^2 + \gamma_{i,1} \rv\gen_i(t) + \gamma_{i,0},
\end{align}
with $\gamma_{i,2} > 0$ for all buses $i \in \mathcal{\nbus}$. However, for a tractable problem we need scalar values in the objective function, not stochastic variables. A common technique is to simply take the expected value. This leads to the new objective function
\begin{align}
\ev{f_i(\rv\gen_i(t))} =  f_i(\ev{\rv\gen_i(t)}) +  \gamma_{i,2}\var{\rv\gen_i(t)} .
\end{align}
\end{subequations}

\subsection{Second-Order Cone Program}

Finally, by combining the results from Sections~\ref{sec:PowerBalance}, \ref{sec:ChanceConstraints}, and \ref{sec:CostFunction}, we present a finite-dimensional and tractable reformulation of Problem~\eqref{eq:CCOPF_original}:
\begin{subequations} 
	\label{eq:SOCP}
	\begin{align}
	\underset{\substack{\hat\gen_i,\, \Gen_{i,j}, \\ \hat\storage_i,\, \Storage_{i,j} \\
	\forall i,j \in \mathcal{\nbus}}}{\operatorname{min}}~  & \sum_{t \in \mathcal{\horizon}}\sum_{i \in \mathcal{\nbus}} f_i(\ev{\rv\gen_i(t)}) +  \gamma_{i,2}\var{\rv\gen_i(t)}  \quad \mathrm{s.\,t.} \\
	\begin{split}
		&\sum_{i \in \mathcal{\nbus}} \hat\load_i + \hat\gen_i + \hat\storage_i = 0_{\horizon}\\
		&\Load_j + \sum_{i \in \mathcal{\nbus}} \Gen_{i,j} + \Storage_{i,j} = 0_{\horizon \times \horizon}, \quad  \forall j \in \mathcal{\nbus}
	\end{split} \\
	& \rv\energy_i(t+1) = \text{\{see Table \ref{tab:FunctionalDepenendencies}\}}, \quad \rv\energy_i(1) = \rv\energy_i^{\text{\textsc{ic}}} \\
	& \underline{x} \leq \ev{\rv{x}} \pm \lambda(\varepsilon) \sqrt{\var{\rv{x}}} \leq \overline{x} \\
	& \sqrt{\var{\rv{x}}} \leq \overline{x}_{\sigma} \\
	\nonumber
	& \forall \rv{x} \in \{ \rv\lineflow_j(t), \, \rv\gen_i(t), \, \Delta\rv\gen_i(\tau), \, \rv\energy_i(t+1), \,\rv\energy_i(\horizon),\, \rv\storage_i(t) \} \\
	\nonumber
	& \forall i \in \mathcal{\nbus}, \, t \in \mathcal{\horizon}, \, \tau \in \mathcal{\horizon} \setminus \{1\}, j \in \mathcal{L}.
	\end{align}
\end{subequations}
Problem~\eqref{eq:SOCP} is a second-order cone program (\socp), hence a convex optimization problem. 

Let us add two more notes on the exact solution and number of decision variables:
As a first note, the \socp provides an \emph{exact} reformulation of Problem~\eqref{eq:CCOPF_original} in the following sense: let $(\rv{\gen}_i(t)^\star, \rv\storage_i(t)^\star)$ for all $i \in \mathcal{N}$ denote the optimal solution to Problem~\eqref{eq:CCOPF_original} restricted to the affine policy~\eqref{eq:AffineFeedback_Gen}, and
let $(\hat{\gen}_i^\star, \Gen_{i,j}^\star, \hat{\storage}_i^\star, \Storage_{i,j}^\star)$ for all $i, j \in \mathcal{\nbus}$ denote the optimal solution to \socp~\eqref{eq:SOCP}.
Applying~\eqref{eq:CC_Reformulation} and~\cite[Proposition 1]{Muehlpfordt17a}, the optimal policies for Problem~\eqref{eq:CCOPF_original} are given by the optimal values of the policy \emph{parameters} via Problem~\eqref{eq:SOCP}
\begin{align}
\begin{bmatrix}
\rv\gen_i(t)^\star \\
\rv\storage_i(t)^\star
\end{bmatrix}
= 
\begin{bmatrix}
[ \hat\gen_i^\star ]_t \\
[ \hat\storage_i^\star ]_t
\end{bmatrix}
+
\sum_{j \in \mathcal{\nbus}} \sum_{k=1}^{t}
\begin{bmatrix}
[   \Gen_{i,j}^\star ]_{tk}   
\\
[   \Storage_{i,j}^\star ]_{tk}
\end{bmatrix}
[\rv\Xi_j]_k
\end{align}
for all buses $i \in \mathcal{\nbus}$ and time instants $t \in \mathcal{\horizon}$.

A second note is that, in theory, the problem is tractable and should be solved efficiently with certified optimality in case of a zero duality gap. However, in practice, large grids may be numerically challenging due to many uncertainties and long horizons $\horizon$. Therefore, it is advisable to introduce a minimum number of scalar decision variables. Specifically, assuming that no bus has both a generator \emph{and} storage, i.e. $\mathcal{\Gen} \cap \mathcal{\Storage} = \emptyset$, for a grid with $N_{\load}$ disturbances, $N_{\gen}$ generators, and $N_{\storage}$ storage systems sets the number of decision variables for local balancing to
\begin{equation}
\label{eq:DecisionVariables_local}
(N_\gen + N_\storage) \left( \horizon + N_\load \frac{\horizon (\horizon + 1)}{2} \right),
\end{equation}
for the generation/storage policies~\eqref{eq:GenerationPolicy}/\eqref{eq:StoragePolicyPerBusPerTime}
\footnote{In contrast to \cite{Warrington13}, we exploit lower-triangularity of the matrices $\Gen_{i,j}$, $\Storage_{i,j}$.}
in local balancing.

In global balancing, see subsection~\ref{sec:localglobal}, for both generation and storage the number of scalar decision variables reduces to
\begin{equation}
\label{eq:DecisionVariables_global}
(N_\gen + N_\storage) \left( \horizon + \frac{\horizon (\horizon + 1)}{2} \right),
\end{equation}
hence it is independent of the number of uncertainties in the grid.
The difference between the numbers~\eqref{eq:DecisionVariables_local} and~\eqref{eq:DecisionVariables_global} reflects the usual trade-off between computational tractability and complexity of the solution.

To summarize: by using affine control policies the infinite-dimensional Problem~\eqref{eq:CCOPF_original} can be written as a tractable convex optimization problem. Since all reformulations are equivalent transformations, there is no loss of information, e.g. all chance constraints from Problem~\eqref{eq:CCOPF_original} are satisfied \emph{exactly}; there is no additional conservatism. Table \ref{tab:comparison_CCOPF_SOCP} illustrates this process.

\begin{table}[ht]
\centering
\caption{Comparison of Problem \eqref{eq:CCOPF_original} and \eqref{eq:SOCP}.}
\label{tab:comparison_CCOPF_SOCP}
\begin{tabular}[t]{lcc}
\toprule
\footnotesize
& Formulation \eqref{eq:CCOPF_original} & Reformulation \eqref{eq:SOCP} \\
\midrule
Problem type    & No SOCP       & SOCP \\
\# constraints  & Infinite      & Finite \\
Solve CCs       & Integral      & Exact formulation \\
Variables       & Random process & Gaussian process \\
Convexity       & Not convex    & Convex \\
Tractability     & \textbf{No}    & \textbf{Yes} \\
\bottomrule
\end{tabular}
\end{table}%
\normalsize

\section{Case Studies}
\label{sec:CaseStudy}

We test the reformulated OPF on various standard test grids of different size over a time horizon of 24h. We start with examining a small network with 5 nodes (\textsc{ieee} case5) in Section \ref{sec:case5} as the solutions are easy to verify and understand. To show that the model works equally well on larger grids, we test the OPF on the 39-bus \textsc{ieee} test case in Section \ref{sec:case39}.
Finally, in Section \ref{sec:complexityAnalysis}, we perform a complexity analysis regarding computation time with the additional grids \textsc{ieee} case57, case118 and case300. 

For all networks, we test three scenarios; without storage (\caseNoStorage), with storage (\caseStorage) and with storage and variance constraints (\caseStorageWithVariance). The variance constraints are introduced by
\begin{align}
\label{eq:VarianceConstraint}
\sqrt{\var{\rv\gen_i(t)}} \leq 0.01.
\end{align}
We test different uncertain disturbances and storage sets, and compare local and global balancing. 
If not stated otherwise, the risk level for each chance constraint in Problem~\eqref{eq:CCOPF_original} is set to $\varepsilon = 5\,\%$ and local balancing is used. In the complexity analysis we use more risk levels ($\epsilon\in\{2.5\%, 5\%, 10\%\}$). 
There are no costs for storage usage; generation costs are the same for all generators. Additionally, storage systems have a prescribed final state, see constraints~\eqref{eq:CCOPF_CCs}, and a maximum capacity.

Apart from showing that the method works well, we answer (i) what importance storage has in a power system with uncertainty, (ii) how scalable our method is in terms of the number of uncertainties and storage, (iii) what influence variance constraints have, (iV) how local and global balancing differ, and (v) what influence different risk levels have.

For the wind forecasts we use a real world wind power data set from ENTSO-E \cite{de_felice_matteo_2021_4682697} that encompasses time series from 2014 to 2021. We smooth the time series with a rolling window of 5 hours and scale according to network capacities. Since the wind farms and data windows are chosen randomly, there is no spatial or temporal correlation that should be considered.

For the sake of simplicity, and without loss of generality, we use the following function to model loads with horizon~$t \in\mathcal{\horizon} = \{1,\dots,24\}$, and, for better understanding, we also use it as a simple, additional forecast for case5:
\begin{subequations}
\label{eq:artificial_forecast}
	\begin{align}
	- [\hat{\load}_i]_t & =  \load_{i}^{\text{nom}} (1 + 0.1 \sin(2 \pi (t-1)/\horizon)), \quad \forall i \in \mathcal{\nbus},\\
	- \Load_i &= 
	\begin{cases}
	\tilde\Load_i \text{ from \eqref{eq:Results_GPVariance_matrix}}, & \forall i \in \mathcal{\Load},\\
	0_{\horizon \times \horizon}, & \forall i \in \mathcal{\nbus} \cap \mathcal{\Load},
	\end{cases}
	\end{align}
\end{subequations}
where~$\load_{i}^{\text{nom}}$ is the nominal load value taken from the case files and $\tilde\Load_i$ is given by \eqref{eq:Results_GPVariance_matrix}.

\begin{figure}
	\centering
	\begin{equation}
	\label{eq:Results_GPVariance_matrix}
	\tilde\Load_i = 10^{-4}\cdot
	\scalemath{0.525}{
		\left[
		\begin{array}{cccccccccccc}
		87   & 0    & 0   & 0   & 0   & 0   & 0   & 0  & 0  & 0  & 0 & 0 \\
		176  & 20   & 0   & 0   & 0   & 0   & 0   & 0  & 0  & 0  & 0 & 0 \\
		292  & 60   & 7   & 0   & 0   & 0   & 0   & 0  & 0  & 0  & 0 & 0 \\
		434  & 124  & 26  & 3   & 0   & 0   & 0   & 0  & 0  & 0  & 0 & 0 \\
		594  & 211  & 63  & 13  & 3   & 0   & 0   & 0  & 0  & 0  & 0 & 0 \\
		764  & 321  & 123 & 31  & 13  & 3   & 0   & 0  & 0  & 0  & 0 & 0 \\
		937  & 447  & 208 & 63  & 32  & 11  & 3   & 0  & 0  & 0  & 0 & 0 \\
		1103 & 582  & 317 & 109 & 65  & 27  & 10  & 3  & 0  & 0  & 0 & 0 \\
		1257 & 718  & 447 & 172 & 116 & 55  & 26  & 10 & 3  & 0  & 0 & 0 \\
		1392 & 847  & 591 & 251 & 184 & 98  & 53  & 26 & 10 & 3  & 0 & 0 \\
		1504 & 964  & 741 & 342 & 271 & 156 & 94  & 53 & 24 & 9  & 3 & 0 \\
		1590 & 1063 & 889 & 441 & 371 & 229 & 151 & 94 & 50 & 24 & 9 & 3
		\end{array}
		\right]
	}
	\end{equation}
	\vspace{\adjustlength}
\end{figure}

For the \gls{gpr} we need to perform a Cholesky decomposition $\Load_i$ of the covariance matrix, to which we apply whitening of $1e^{-7}$ due to slight numerical instabilities.
\gls{gpr} was implemented in Python \cite{10.5555/1593511} version 3.8.8 using GpFlow \cite{GPflow2017} based on tensorflow.
The SOCPs were implemented in Julia~\cite{Bezanson2017} version 1.6.1, and solved with \textsc{j}u\textsc{mp}~\cite{Dunning2017} and the MOSEK solver set to its default values, using a PC with an AMD Ryzen™ 7 PRO 4750U processor at 1700 Mhz and 16GB memory \cite{muhlpfordt_git_nodate}.

\subsection{\textsc{ieee} 5-bus test case}
\label{sec:case5}

\begin{figure}
	\begin{subfigure}[c]{\figwidth}
    \centering
    \includegraphics[width=0.95\textwidth]{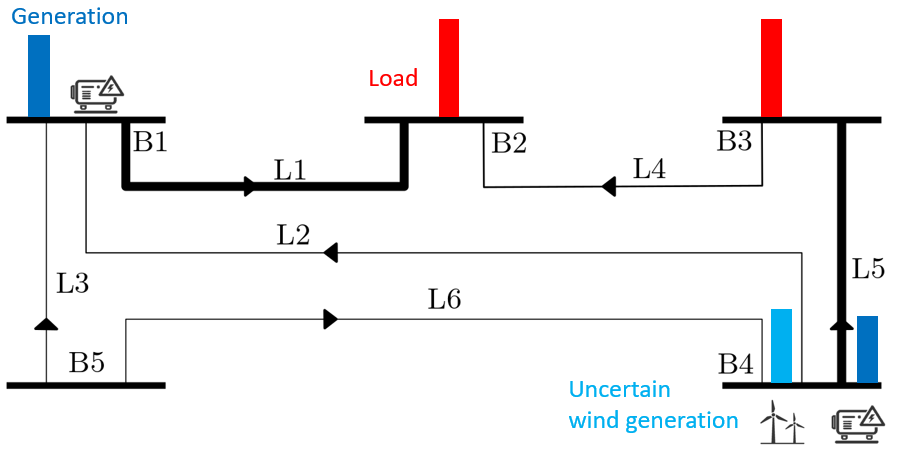}
    \caption{Without storage.}
    \label{fig:case5_results_grids_NoStorage}
\end{subfigure}
\begin{subfigure}[c]{\figwidth}
    \centering
    \includegraphics[width=0.95\textwidth]{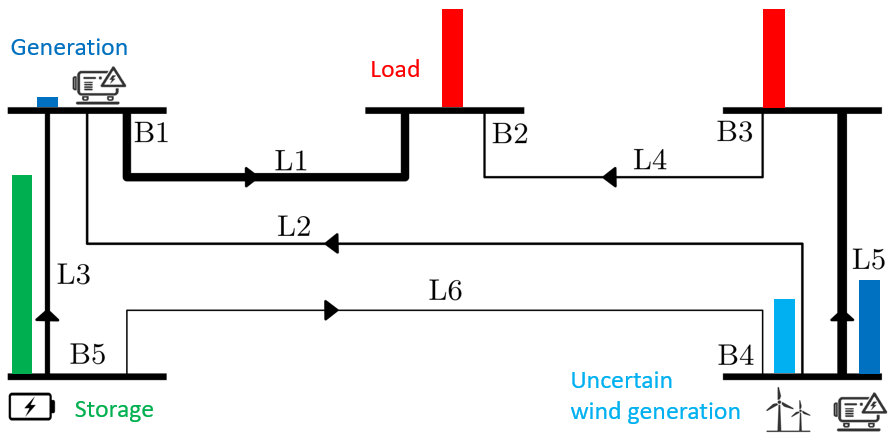}
    \caption{With storage.}
    \label{fig:case5_results_grids_Storage}
    \end{subfigure}
    \caption{\textsc{ieee} case5: Network state without (upper) and with (lower) storage at time $t=3$, with generation (dark blue), wind generation (light blue), loads (red), storage (green) and line flows (black).}
    \label{fig:case5:results_grids}
\end{figure}

Let us first apply method \eqref{eq:SOCP} to a simple test network in order to foster a good understanding of the dynamics. 
\textsc{ieee} case5 has five nodes, six lines, and two generators at buses $\mathcal{\Gen} = \{1,4\}$. We install two loads at buses $\{2,3\}$, one storage at bus $\mathcal{\Storage} = \{5\}$ and one uncertain disturbance at bus $\mathcal{\Load} = \{4\}$ that represents a wind farm, see Figure \ref{fig:case5:results_grids}. 

We alter the case file slightly in order to make it compatible with our method: Generators 1 and 2 are merged (by adding their capacities Pg, Qg, Qmax, Qmin, Pmax), because the program requires maximal one generator per node. 
And generator 5 is replaced by a storage, as each node can only contain a generator \emph{or} a storage. 
All minor changes, such as cost coefficients and line ratings, can be found in Table \ref{tab:Parameters}.

\begin{figure}
    \centering
    \includegraphics[width=0.5\textwidth]{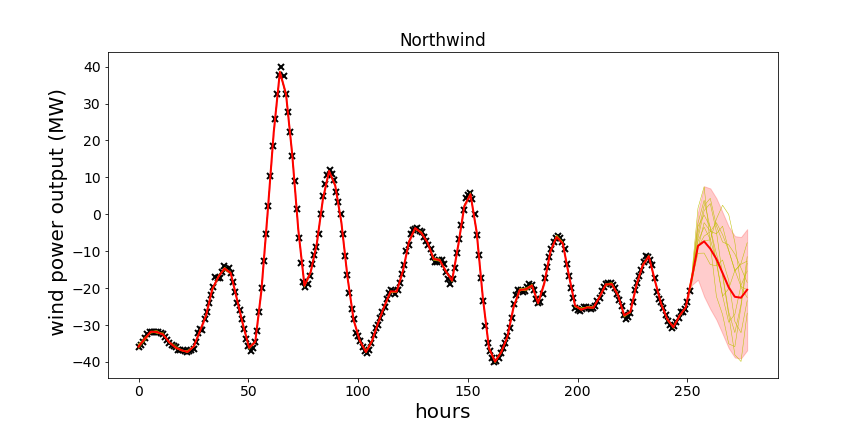}
    \caption{GPR-fitted and forecast wind power outputs smoothed with a rolling window of 5.}
    \label{fig:wind_forecast_volatile}
\end{figure}

Besides the network, the OPF requires a second input; wind forecast in the form of  \glspl{gp}.
Figure \ref{fig:wind_forecast_volatile} shows the forecast of wind power for a random day of wind farm \emph{Northwind}. We selected the kernel as in equation \eqref{eq:kernel}. 
As we can see, the GPR fits the given data well, while the horizon encompasses more variance (uncertainty).

\begin{figure}
    \begin{subfigure}[c]{\figwidth}
        \centering
        \includegraphics[width=0.45\textwidth]{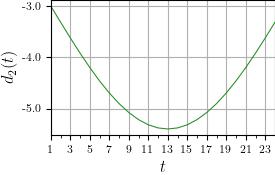}
        \includegraphics[width=0.45\textwidth]{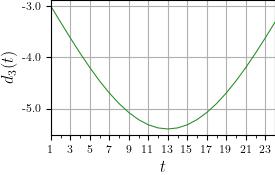}
        \includegraphics[width=0.6\textwidth, height=0.3\textwidth]{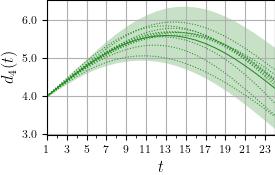}
        \vspace{-2mm}
        \caption{Upper: certain disturbances~$-\load_{2}(t), \load_{3}(t)$ at buses $\{2,3\}$; lower: ten realizations of the uncertain disturbance~$-\rv\load_4(t)$ at bus~$4$ (dots), mean $-\ev{\rv\load_4(t)}$ (solid), and $-(\ev{\rv\load_4(t)} \pm 3\sqrt{\var{\rv\load_4(t)}})$-interval (shaded).}
        \label{fig:case5:Loads}
    \end{subfigure}
    
	\begin{subfigure}[c]{\figwidth}
		\centering
		
        \includegraphics[width=0.45\textwidth]{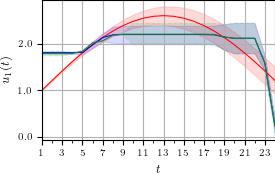}~
        \includegraphics[width=0.45\textwidth]{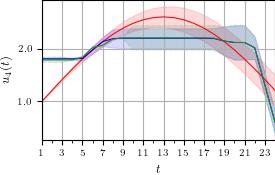}%
		
        \includegraphics[width=0.45\textwidth]{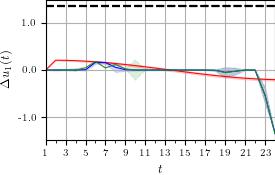}~
        \includegraphics[width=0.45\textwidth]{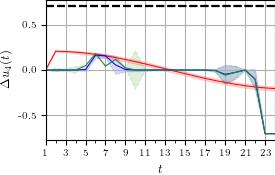}%
        
		\vspace{-2mm}		
		\caption{Upper: Power injections of generator at buses $\{1, 4\}$; lower: respective change in power injections.}
		\label{fig:case5:Generation}
	\end{subfigure}
	
	\begin{subfigure}[c]{\figwidth}
		\centering
        \includegraphics[width=0.45\textwidth]{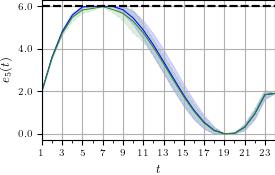}~
        \includegraphics[width=0.45\textwidth]{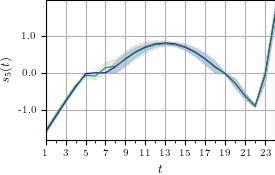}
		
		\vspace{-2mm}
		\caption{Left: Power injections of storage at bus $5$; right: respective change of power.}
		\label{fig:case5:Storage}
	\end{subfigure}
	\begin{subfigure}[c]{\figwidth}
	\centering
        \includegraphics[width=0.45\textwidth]{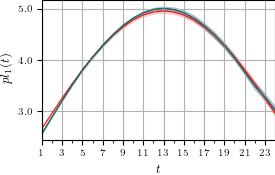}
        \includegraphics[width=0.45\textwidth]{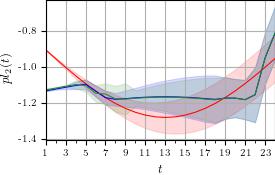}
        \includegraphics[width=0.45\textwidth]{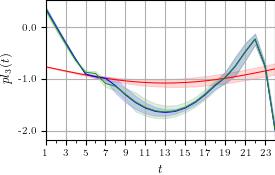}
        \includegraphics[width=0.45\textwidth]{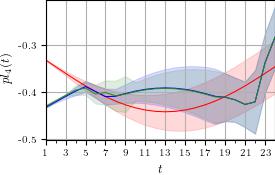}
        \includegraphics[width=0.45\textwidth]{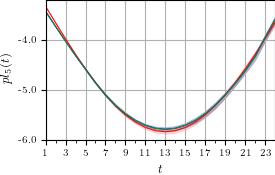}
        \includegraphics[width=0.45\textwidth]{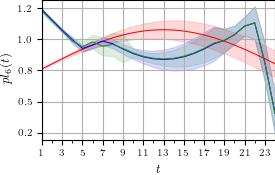}
		\vspace{-2mm}		
		\caption{Line flows across all lines.}
		\label{fig:case5:LineFlows}
	\end{subfigure}
	\vspace{\adjustlength}
	\caption{\textsc{ieee} 5-bus grid with artificial wind generation: Results for cases \caseNoStorage (red), \caseStorage (blue), and \caseStorageWithVariance (green). All shown random variables~$\rv{x}$ are depicted in terms of their mean~$\ev{\rv{x}}$ (solid) and the interval $\ev{\rv{x}} \pm \lambda (0.05) \sqrt{\var{\rv{x}}}$ (shaded).}
	\label{fig:case5:results_artificial_newLoads}
\end{figure}

\begin{figure}
    \begin{subfigure}[c]{\figwidth}
        \centering
        \includegraphics[width=0.45\textwidth]{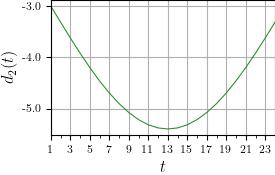}
        \includegraphics[width=0.45\textwidth]{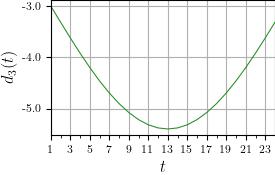}
        \includegraphics[width=0.6\textwidth, height=0.3\textwidth]{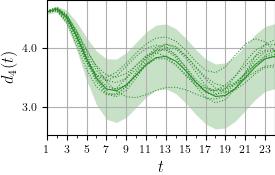}
        \vspace{-2mm}
        \caption{Upper: certain disturbances~$-\load_{2}(t), \load_{3}(t)$ at buses $\{2,3\}$; lower: ten realizations of the uncertain disturbance~$-\rv\load_4(t)$ at bus~$4$ (dots), mean $-\ev{\rv\load_4(t)}$ (solid), and $-(\ev{\rv\load_4(t)} \pm 3\sqrt{\var{\rv\load_4(t)}})$-interval (shaded).}
        \label{fig:case5:Loads_volatile}
    \end{subfigure}
	\begin{subfigure}[c]{\figwidth}
		\centering
        \includegraphics[width=0.45\textwidth]{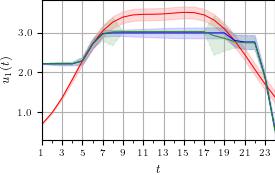}~
        \includegraphics[width=0.45\textwidth]{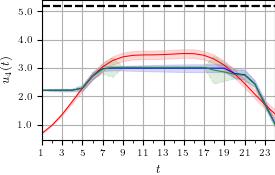}%
		
        \includegraphics[width=0.45\textwidth]{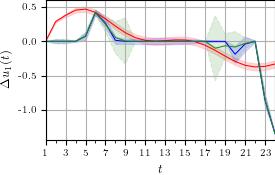}~
        \includegraphics[width=0.45\textwidth]{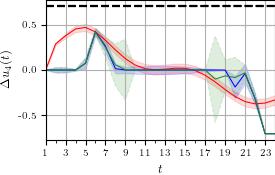}%
		\vspace{-2mm}		
		\caption{Upper: Power injections of generator at buses $\{1, 4\}$; lower: respective change in power injections.}
		\label{fig:case5:Generation_volatile}
	\end{subfigure}
	
	\begin{subfigure}[c]{\figwidth}
		\centering
        \includegraphics[width=0.45\textwidth]{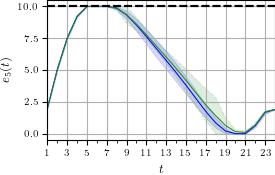}~
        \includegraphics[width=0.45\textwidth]{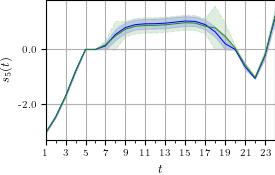}
		
		\vspace{-2mm}
		\caption{Left: Power injections of storage at bus $5$; right: respective change of power.}
		\label{fig:case5:Storage_volatile}
	\end{subfigure}
	\begin{subfigure}[c]{\figwidth}
	\centering
        \includegraphics[width=0.45\textwidth]{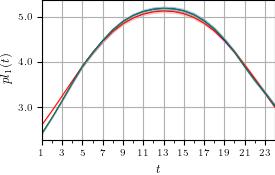}
        \includegraphics[width=0.45\textwidth]{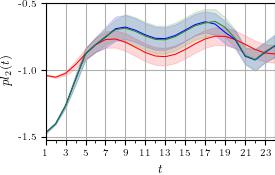}
        \includegraphics[width=0.45\textwidth]{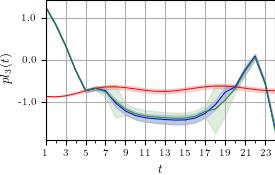}
        \includegraphics[width=0.45\textwidth]{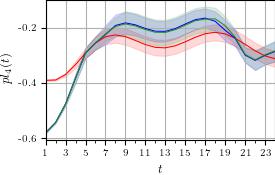}
        \includegraphics[width=0.45\textwidth]{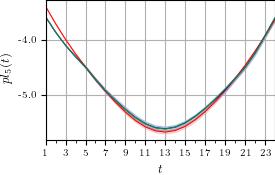}
        \includegraphics[width=0.45\textwidth]{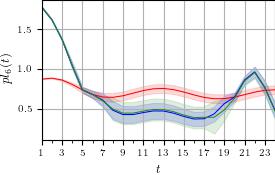}
		\vspace{-2mm}		
		\caption{Line flows across all lines.}
		\label{fig:case5:LineFlows_volatile}
	\end{subfigure}
	\vspace{\adjustlength}
	\caption{\textsc{ieee} 5-bus grid with real-world wind generation: Results for cases \caseNoStorage (red), \caseStorage (blue), and \caseStorageWithVariance (green). All shown random variables~$\rv{x}$ are depicted in terms of their mean~$\ev{\rv{x}}$ (solid) and the interval $\ev{\rv{x}} \pm \lambda (0.05) \sqrt{\var{\rv{x}}}$ (shaded).}
	\label{fig:case5:results_volatile_newLoads}
\end{figure}

The OPF results for the predicted horizon with artificial and real-world forecasts are given by Figures \ref{fig:case5:results_artificial_newLoads}, that we describe in detail, and by Figure \ref{fig:case5:results_volatile_newLoads}, that works analogously.
Generation, storage and line images contain several coloured curves that depict the different scenarios; without storage (red), with storage (blue), and storage with variance constraints on generators (green).
Figure \ref{fig:case5:Loads} shows the loads and ten realizations of the uncertain wind generation. Note how the variance grows slightly over time and then stops to grow in the later third. This behaviour seems realistic as uncertainty grows early in time, but even forecasts cannot take on unlimited uncertainty.

Generation and change in generation is given in Figure \ref{fig:case5:Generation}. Without storage (red), the generator needs to provide the difference in power between demand and wind generation. Hence, it reflects the behaviour of the sum of load and wind generation (in this case they have the same behaviour), and assumes all uncertainty of the forecast. 
In contrast, in the scenarios with storage \caseStorage (blue) and additional variance constraints \caseStorageWithVariance (green), the generation curves are a lot more constant, and assume a constant level of variance. Looking closely, the variance constraint for \caseStorageWithVariance almost diminishes variance for times $t=3,\dots,9$.
At the end of the horizon, generation curves go down as they have to respond with final storage constraints. 

Storage is depicted in Figure \ref{fig:case5:Storage}. Since there is a surplus of wind generation up to $t=4$, the storage is filled to its limit. Afterwards, the load surpasses generation and the storage empties. Much of the variance is absorbed by the storage;  even more so in scenario \caseStorageWithVariance due to the variance restriction of the generator.

Line flows of all six transmission lines are shown in Figure \ref{fig:case5:LineFlows}. Most obviously, they mirror the loads and uncertain wind generation. 
Without storage, all lines mirror the sum of load and wind generation. 
Upon including storage, lines $1$ and $5$ still mirror the load as they directly connect a generator with a load (see Figure \ref{fig:case5:results_grids}). The other lines are slightly smoothed as they are influenced by the storage.

Replacing the artificial wind forecast with a GPR prediction on real-world data introduces volatility (see Figure \ref{fig:case5:Loads_volatile}).
This leads to a lot more fluctuation for the generators with no storage (see Figure \ref{fig:case5:Generation_volatile}). Including storage leads again to almost constant generation, however, we have to increase the storage capacity to $10$ for a similar effect. In terms of storage and line flow there are no differences; the OPF works alike in both trials (see Figures \ref{fig:case5:Storage_volatile} and \ref{fig:case5:LineFlows_volatile}).

Figure \ref{fig:case5:results_grids} visualizes the grids mean values at point $t=3$ in time, for the artificial load, without and with storage (\caseNoStorage and \caseStorage). At this point in time, storage is fully charged and the effect it has on the grids dynamics becomes clearest. 
Figure \ref{fig:case5_results_grids_NoStorage} shows \caseNoStorage, while Figure \ref{fig:case5_results_grids_Storage} shows \caseStorage.
The effect of storage is that it drastically reduces generation, despite high load.

\begin{figure}
	\centering
	\vspace{3cm}
	
	\begin{subfigure}[c]{\figwidth}
		\centering
		\includegraphics[width=0.45\textwidth]{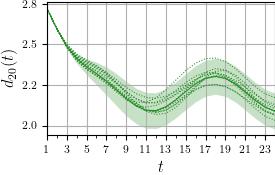}~
		\includegraphics[width=0.45\textwidth]{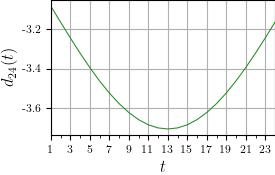}%
		
		\vspace{-2mm}		
		\caption{Left: ten realizations of the uncertain disturbance~$-\rv\load_20(t)$ at bus~$20$ (dots), mean $-\ev{\rv\load_20(t)}$ (solid), and $-(\ev{\rv\load_20(t)} \pm 3\sqrt{\var{\rv\load_20(t)}})$-interval (shaded); right: certain disturbance~$-\load_{24}(t)$ at bus $3$.}
		\label{fig:Disturbances}
	\end{subfigure}
	
	\begin{subfigure}[c]{\figwidth}
		\centering
        \includegraphics[width=0.45\textwidth]{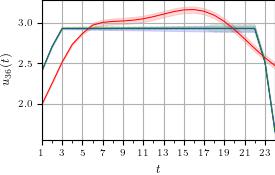}~
        \includegraphics[width=0.45\textwidth]{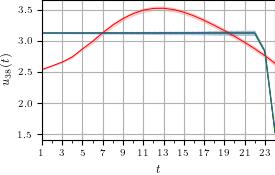}%
		
        \includegraphics[width=0.45\textwidth]{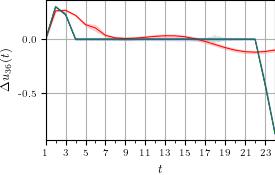}~
        \includegraphics[width=0.45\textwidth]{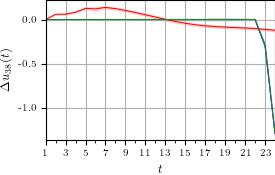}%
        
		\vspace{-2mm}		
		\caption{Upper: Power injections of generators at buses $i \in \{36,38\}$ ; lower: respective change in power injections.}
		\label{fig:Generation}
	\end{subfigure}
	
	\begin{subfigure}[c]{\figwidth}
		\centering
        \includegraphics[width=0.45\textwidth]{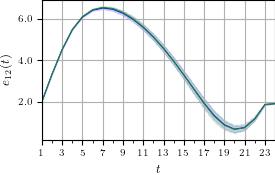}~
        \includegraphics[width=0.45\textwidth]{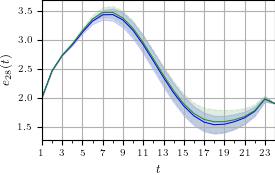}%
        
		\vspace{-2mm}
		\caption{Upper: Power injections of storages at buses $i \in \{2,5\}$; lower: respective state of storage.}
		\label{fig:Storage}
	\end{subfigure}
	
	\begin{subfigure}[c]{\figwidth}
	\centering
        \includegraphics[width=0.45\textwidth]{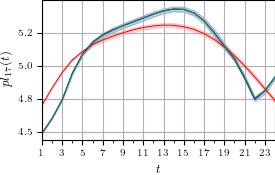}~
        \includegraphics[width=0.45\textwidth]{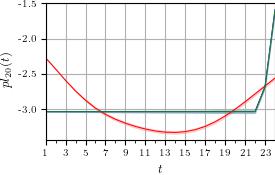}%
        
		\vspace{-2mm}		
		\caption{Line flows across lines $l \in \{17,20\}$.}
		\label{fig:LineFlows}
	\end{subfigure}
	
	\vspace{\adjustlength}
	\caption{\textsc{ieee} 39-bus grid: Results for cases \caseNoStorage (red), \caseStorage (blue), and \caseStorageWithVariance (green). All shown random variables~$\rv{x}$ are depicted in terms of their mean~$\ev{\rv{x}}$ (solid) and the interval $\ev{\rv{x}} \pm \lambda (0.05) \sqrt{\var{\rv{x}}}$ (shaded).}
	\label{fig:IEEEresults}
	\vspace{3cm}
\end{figure}

\subsection{\textsc{ieee} 39-bus test case}
\label{sec:case39}

\begin{figure}
    \centering
    \includegraphics[width=0.5\textwidth]{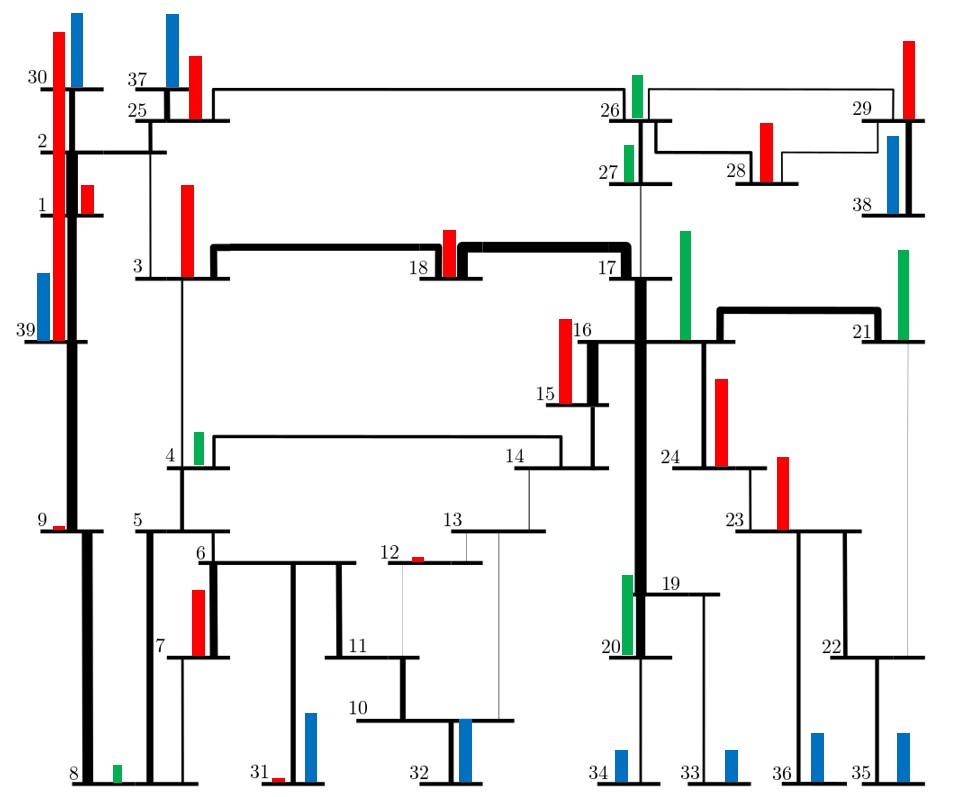}
    \caption{\textsc{ieee} case39: Network state without (upper) and with (lower) storage at time $t=9$, with generation (dark blue), wind generation (light blue), loads (red), storage (green) and line flows (black).}
    \label{fig:case39_grid}
\end{figure}

After having tested method \eqref{eq:SOCP} on a small grid, we show that it works equally well on a larger grid.
The \textsc{ieee} 39-bus system has a total of 10 generators and 46 lines~\cite{Zimmerman11}, see Figure \ref{fig:case39_grid}.
We introduce seven uncertain disturbances at buses~$\mathcal{\Load} = \{4,8,16,20,21,26,27\}$, and five storages are placed at buses~$\mathcal{\Storage} =  \{1,12,14,18,28\}$.
Table~\ref{tab:Parameters} in the Appendix collects all problem-relevant parameters; storage is at 10 again.

In order to check the method and see that storages have the same effect as before, we look at the optimized horizon~$\mathcal{\horizon} = \{1,\dots,24\}$ in Figure \ref{fig:IEEEresults}. The plots are fairly representative for the grid, i.e. the other components behave alike.
Load and wind generation, Figure \ref{fig:Disturbances}, only differ in size, as they are adjusted to the network parameters.
Generation, storage and line flow curves behave similarly.
More components are given in \ref{app:plots}: other loads are equivalent; remaining generators, storages and line flows behave similarly.
Hence, the method also works on this larger grid.

Figure \ref{fig:case39_grid} depicts the grid with all components and line flows. We can see that at time $t=9$ storages are filled and lines adjacent to storage are loaded heavily. Generation is less than in scenario \caseNoStorage without storage.



\subsection{Computational complexity}
\label{sec:complexityAnalysis}

To evaluate the method in terms of scalability, we add \textsc{ieee} cases case57, case118 and case300 to the previous two and perform a complexity analysis with regard to computation time and costs.
Uncertainties are placed at the nodes with the highest load, i.e. the highest impact, and storage systems are placed randomly as placement does not influence computation time. 
We analyse the role of the network size, of the number of uncertain disturbances, of local vs. global balancing, and of storage on computation time. 
Additionally, we show how the costs differ with respect to risk levels, global vs. local balancing as defined in Section \ref{sec:localglobal} and storage. 
\begin{figure}
    \centering
    \includegraphics[width=0.50\textwidth]{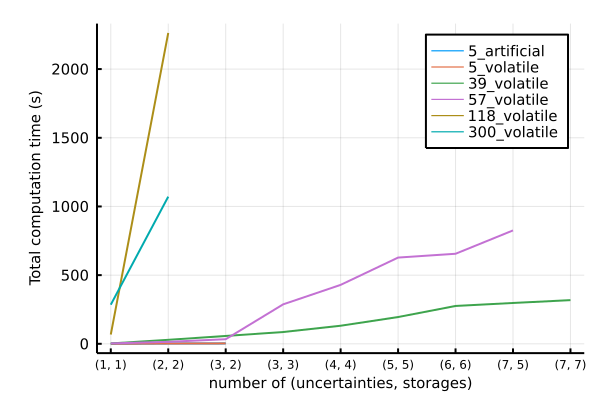}
    \caption{Computation time of all test cases with respect to the number of uncertainties and storages.}
    \label{fig:complexity_all}
\end{figure}

Figure \ref{fig:complexity_all} shows the computational complexity for all cases with one to eight uncertain loads and storage installations. While smaller cases run within seconds to a couple of minutes, the run time for larger network sizes above 57 rapidly increases to more than 15 minutes. We can compute up to 118 nodes; for a larger number of nodes Mosek runs out of space. Hence, the number of nodes drives computation time up considerably. Concerning the time horizon, we have compared the horizons with 12 and 24 time steps. Since the time horizon enters quadratically into the number of equations, it does have significant influence; about factor 10 in computation time.

\begin{figure}
    \centering
    \includegraphics[width=0.55\textwidth]{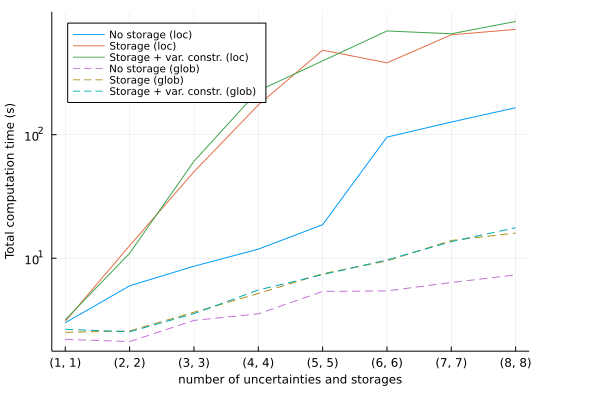}
    \caption{Computation time for case57 of each scenario for local and global balancing with respect to the number of uncertainties and storage (log scale).}
    \label{fig:57_complexity_szenarios}
\end{figure}

We compare the role of different scenarios and local vs. global balancing with the example of case57, in Figure \ref{fig:57_complexity_szenarios}. Clearly, local balancing takes a lot longer than global balancing. Also, storage increases computation time significantly, while adding variance constraints does not, as expected. The number of decision variables (blue points) scales linearly with the number of uncertainties plus storages, as can be seen from equation \eqref{eq:DecisionVariables_local}.
Other cases behave similarly.

\begin{figure}
    \includegraphics[width=0.55\textwidth]{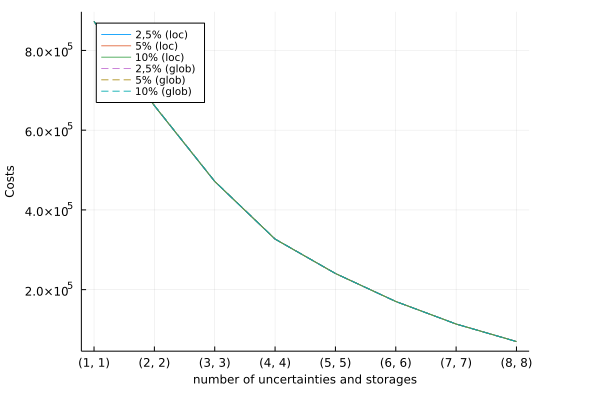}
    \caption{\textsc{ieee} case39: Costs with respect to the different scenarios \caseNoStorage, \caseStorage and \caseStorageWithVariance, different risk levels $\epsilon\in\{2.5\%, 5\%, 10\%\}$ and local vs. global balancing.}
    \label{fig:39_complexity_risk_level}
\end{figure}

\begin{figure}
    \includegraphics[width=0.45\textwidth]{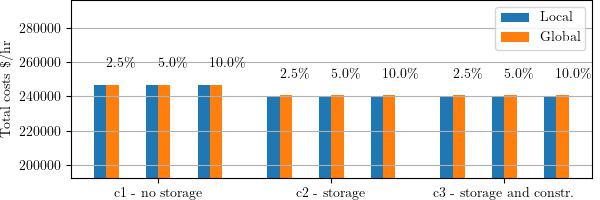}
    \caption{\textsc{ieee} case39 with 5 uncertainties and storages: Costs with respect to the different scenarios \caseNoStorage, \caseStorage and \caseStorageWithVariance, different risk levels $\epsilon\in\{2.5\%, 5\%, 10\%\}$ and local vs. global balancing.}
    \label{fig:39_costs_bars}
\end{figure}

Another interesting measure is cost. In our cases costs do not differ largely, partly because the costs themselves are high relative to the differences. 
Figure \ref{fig:39_complexity_risk_level} shows that costs fall with a higher number of uncertainties and storages. This could be explained by a larger balancing effect and less costs through conventional generation.

Figure \ref{fig:39_costs_bars} shows in more detail the differences for the cases and local and global optimization for case39 with 5 uncertainties and storages. We can see that using no storage (\caseNoStorage) has higher cost than using storage, which holds for the other cases, too. Also, global optimization is at least as expensive as local balancing. The risk levels hardly influence costs.

\section{Discussion}
\label{sec:discussion}

The main result from Sections \ref{sec:case5} and \ref{sec:case39} is that the method works equally well on various network sizes.
Moreover, we show three outcomes:
(i) Generation profiles are flattened out, hence, generation is a lot more stable with storage in use.
(ii) Costs reduce when more storage and uncertainties are in use, and generation and storage profiles are more similar. This suggests that larger networks can balance out uncertainties better, hence, they are more stable and secure.
(iii) Most of the uncertainty in the wind forecast is absorbed by storage, which means that renewable energy can be well integrated into generation planning, even if there is a lot of uncertainty.

Adding a remark about convergence, we can tell that the network does not converge in two cases: Firstly, when demand is larger than generation, as expected. And secondly, also as expected, when demand is too high in the beginning, because generators cannot ramp up fast enough as they reach their ramp limits.

From Section \ref{sec:complexityAnalysis}, testing computation time and costs, we can derive five results:
(i) The method is scalable up to roughly 100 nodes without any speedup (e.g. sparsity methods, contraction algorithms).
(ii) Risk levels do not influence costs or computation time.
(iii) local balancing takes a lot longer than global balancing, nevertheless reduces the costs slightly.
(iv) Computation time with respect to the number of uncertainties does faster than linearly with the number of decision variables, that is the number of uncertainties, storages, and the time horizon.
(v) Storages reduce generation costs notably.
Hence, the method works well on mid-size power grids and is fairly robust with respect to parameter variations. 

Concluding, we can say that the method is robust and performs well on mid-size networks. However, matrix sparcity and contraction algorithms offer large potential for speed-up. Additionally, storage plays a large role in cost reduction, reducing uncertainty by renewables, and stabilizing generation.


\section{Conclusions and Outlook}
\label{sec:conclusion}

\glsreset{gp}
\glsreset{gpr}

We reformulate an intractable DC optimal power flow problem with uncertain disturbances and chance constraints into a tractable second order cone problem with exact analytical expressions. We model all decision variables as \glspl{gp} and predicted the disturbances with \gls{gpr}.
We test the approach on networks of differing sizes.
The new problem formulation with \glspl{gp} capturing uncertainty gives realistic results and is computationally efficient for mid-size networks.
The model shows that uncertainty can be handled well by including storage into transmission networks. Almost all uncertainty is absorbed and little left at the generators, which allows for stable generation scheduling. 
Without storage much uncertainty is left at the generators and network control becomes a much more difficult and uncertain task.
Including storage also reduces the cost notably, even with variance constraints.

Further research should aim to adapt the method for practical use. 
As real-world networks are often very large, speeding up the algorithm is a next goal, for example by using the sparsity of matrices.
Also, one can look at non-Gaussian disturbances, or give more detail to the modelling of generators and storage. 
An interesting part will be to automate the \gls{gpr}  with large amounts of data.


\section{Authors contribution and acknowledgements}

\textbf{Rebecca Bauer (shared first author):} Data Curation, Software: GPR and parts of SOCP, Analysis, Writing: Original Draft, Review \& Editing, Visualization; 
\textbf{Tillmann Mühlpfordt (shared first author):} Conceptualization, Methodology, Software: SOCP, Validation, Analysis, Writing: Original Draft, Visualization, Supervision, Project administration;
\textbf{Nicole Ludwig:} Supervision, Conceptualization, Writing: Original Draft, Review \& Editing;
\textbf{Veit Hagenmeyer:} Supervision, Conceptualization, Review \& Editing, Project administration, Funding acquisition


Rebecca Bauer acknowledges funding by the BMBF-project MOReNet with grant number 05M18CKA.

Nicole Ludwig acknowledges funding by the Deutsche Forschungsgemeinschaft (DFG, German Research Foundation) under Germany’s Excellence Strategy – EXC number 2064/1 – Project number 390727645 and the Athene Grant of the University of Tübingen.


\bibliography{article}


\clearpage

\appendix

\section{Parameter values for case studies}
\label{app:parameters}

\begin{table}[h]
	\centering
	\resizebox{0.85\linewidth}{!}{%
		\centering
		\begin{minipage}{\linewidth}
			\centering
			\caption{Parameter values for case studies.}	
			\label{tab:Parameters}
			\centering	
			\begin{tabular}{p{0.45cm}llll}	
				\toprule
				\multirow{2}{2.75em}{$i \in \mathcal{\Gen}$}   &   $\underline{\gen}_i=0.0$ &   $\overline{\gen}_i =\,1.1 \overline{p}_i$ &       $\Delta \underline{\gen}_i =-0.15\overline{p}_i$ &        $\Delta \overline{\gen}_i =0.15\overline{p}_i$\\
				&      $\gamma_{i,2} = 0.01$ &      $\gamma_{i,1}=0.3$ &              $\gamma_{i,0} =0.2$ &  \\
				\midrule
				\multirow{2}{2.75em}{$i \in \mathcal{\Storage}$} &  $\underline{\energy}_i =0.0$ &  $
				\overline{\energy}_i =6.0$ &         $\underline{\storage}_i  =-10.0$ &          $\overline{\storage}_i = 10.0$\\
				& $\underline{\energy}_i^\horizon = 0.19 $ & $\overline{\energy}_i^\horizon =0.21 $ & $\ev{\rv\energy_i^{\text{\textsc{ic}}}} =2.0 $ & $\var{\rv\energy_i^{\text{\textsc{ic}}}} = 0.0$\\
				\midrule
				\multirow{1}{2.75em}{$j \in \mathcal{L}$}     & $\underline{\lineflow}_j=-0.85\overline{p}_{l,j}$ & $\overline{\lineflow}_j = 0.85 \overline{p}_{l,j}$ &  \multicolumn{2}{|c}{$\overline{p}_i$, $\overline{p}_{l,j}$ taken from case file \cite{Zimmerman11}}\\ \bottomrule
			\end{tabular}
		\end{minipage}
	}
	\vspace{\adjustlength}
\end{table}

\vspace{5mm}

\clearpage

\section{Additional plots of case studies}
\label{app:plots}






\begin{figure}[h]
	\centering
    
    \begin{subfigure}[c]{\figwidth}
    \centering
        \includegraphics[width=0.45\textwidth]{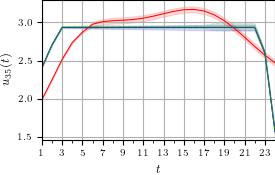}~
    	\includegraphics[width=0.45\textwidth]{figures/time series/case39_volatile_24/gen_u_8109.jpg}%
    	\vspace{-2mm}	
    	\caption{Power injections of generators at buses $i \in \{6,9\}$.}
    \end{subfigure}
    
    \begin{subfigure}[c]{\figwidth}
    \centering
    	\includegraphics[width=0.45\textwidth]{figures/time series/case39_volatile_24/storage_e_8652.jpg}~
    	\includegraphics[width=0.45\textwidth]{figures/time series/case39_volatile_24/storage_e_8655.jpg}%
    	\vspace{-2mm}	
    	\caption{Power injections of storages at buses $i \in \{2,5\}$.}
    \end{subfigure}
    
    \begin{subfigure}[c]{\figwidth}
    \centering
        \includegraphics[width=0.45\textwidth]{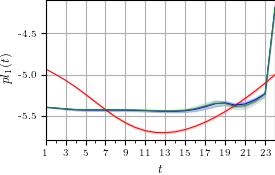}~
    	\includegraphics[width=0.45\textwidth]{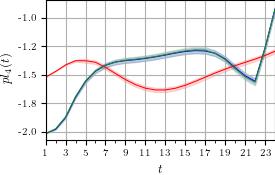}%
    \end{subfigure}
    
    \begin{subfigure}[c]{\figwidth}
    \centering
    	\includegraphics[width=0.45\textwidth]{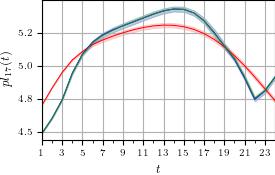}~
    	\includegraphics[width=0.45\textwidth]{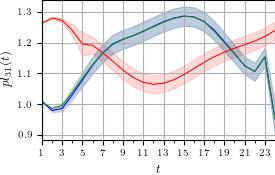}%
    	\vspace{-2mm}	
    	\caption{Power flows at lines $i\in\{1,4,17,31\}$.}
    \end{subfigure}
    	
    \vspace{2mm}	
	\vspace{\adjustlength}
	\caption{\textsc{ieee} 39-bus grid: Results for 7 uncertainties and 5 storage systems for cases \caseNoStorage (red), \caseStorage (blue), and \caseStorageWithVariance (green). All shown random variables~$\rv{x}$ are depicted in terms of their mean~$\ev{\rv{x}}$ (solid) and the interval $\ev{\rv{x}} \pm \lambda (0.05) \sqrt{\var{\rv{x}}}$ (shaded).}
\end{figure}


\begin{figure}
	\centering
    
    \begin{subfigure}[c]{\figwidth}
    \centering
        \includegraphics[width=0.45\textwidth]{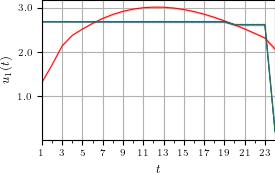}~
    	\includegraphics[width=0.45\textwidth]{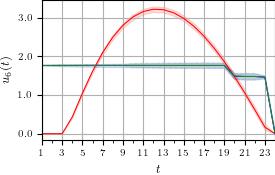}%
    	\vspace{-2mm}	
    	\caption{Power injections of generators at buses $i \in \{1,4\}$.}
    \end{subfigure}
    
    \begin{subfigure}[c]{\figwidth}
    \centering
    	\includegraphics[width=0.45\textwidth]{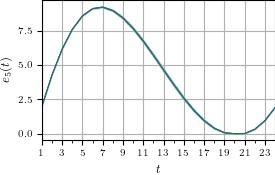}~
    	\includegraphics[width=0.45\textwidth]{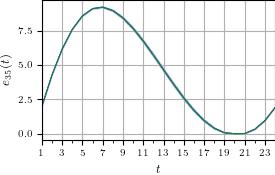}%
    	\vspace{-2mm}	
    	\caption{Power injections of storages at buses $i \in \{1,4\}$.}
    \end{subfigure}
    
    \begin{subfigure}[c]{\figwidth}
    \centering
        \includegraphics[width=0.45\textwidth]{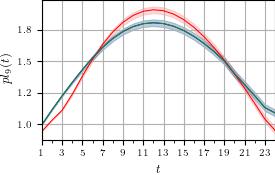}~
    	\includegraphics[width=0.45\textwidth]{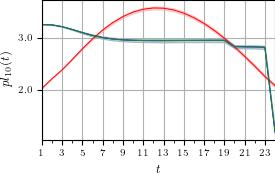}%
    \end{subfigure}
    \begin{subfigure}[c]{\figwidth}
    \centering
    	\includegraphics[width=0.45\textwidth]{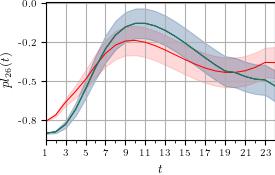}~
    	\includegraphics[width=0.45\textwidth]{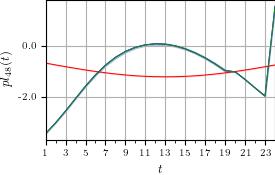}%
    	\vspace{-2mm}	
    	\caption{Power flows at lines $i\in\{9,10,26,48\}$.}
    \end{subfigure}
    	
    \vspace{2mm}	
	\vspace{\adjustlength}
	\caption{\textsc{ieee} 57-bus grid: Results for 7 uncertainties and 5 storage systems for cases \caseNoStorage (red), \caseStorage (blue), and \caseStorageWithVariance (green). All shown random variables~$\rv{x}$ are depicted in terms of their mean~$\ev{\rv{x}}$ (solid) and the interval $\ev{\rv{x}} \pm \lambda (0.05) \sqrt{\var{\rv{x}}}$ (shaded).}
\end{figure}

\end{document}